\documentclass[%
 aip,
 amsmath,amssymb,
reprint, 
onecolumn 
]{revtex4-1}

\usepackage{graphicx}
\usepackage{dcolumn}
\usepackage{bm}
\usepackage[mathlines]{lineno}
\newcommand{\bi}{\bibitem}
\usepackage[utf8]{inputenc}
\usepackage[T1]{fontenc}
\usepackage{mathptmx}
\usepackage{etoolbox}
\usepackage{amsmath}
\usepackage{amssymb}
\usepackage[colorlinks, citecolor=blue, linkcolor=red, urlcolor=blue]{hyperref}
\usepackage{amscd}
\usepackage{lipsum}
\usepackage{xcolor}
\usepackage[colorlinks]{hyperref}

\usepackage{etoolbox}

\makeatletter
\def\@email#1#2{%
 \endgroup
 \patchcmd{\titleblock@produce}
  {\frontmatter@RRAPformat}
  {\frontmatter@RRAPformat{\produce@RRAP{*#1\href{mailto:#2}{#2}}}\frontmatter@RRAPformat}
  {}{}
}%
\makeatother
\begin{document}

\title{Fixed-Point Abundances and the Onset of Extinctions in Multispecies Lotka-Volterra Ecosystems}
\author{Ph. Jacquod}
\email{philippe.jacquod@unige.ch}
\affiliation{Department of Quantum Matter Physics, University of Geneva, Geneva, Switzerland
}%
 \affiliation{School of Engineering, University of Applied Sciences of Western Switzerland, CH-1950 Sion, Switzerland}

\date{\today}

\begin{abstract}
I investigate the average and variance of  species abundance distributions for fixed-point solutions to generalized Lotka-Volterra ecosystems. 
Two different analytical methods are constructed. First, I build up on the perturbative Random Matrix Theory approach developed in my earlier work [Phys. Rev. E. {\bf 113}, L062202 (2026)],
and formalize this method diagrammatically. The abundance average is given by a sum of planar diagrams with non-crossing vertex pairings which can 
be encoded as Dyck paths. This allows me to express the combinatorial factor multiplying perturbative terms at each order as Catalan and Narayana numbers, following which 
the abundance average is given by the latter's generating functions. In this way, the infinite perturbation expansion is exactly resummed in a closed-form formula. I have not been able to
find a similar resummation for the abundance variance, however diagrammatic Feynman rules are expressed, following which the perturbative series can be calculated to 
any desirable order. Second, I do the same calculations using the cavity method, which gives closed-form formulas for  both the abundance average and variance. 
For the average, the cavity formula gives the exact same result as the perturbative approach when the average interspecies interaction vanishes. 
When this is not the case, the two results are functionally different, however they are hardly distinguishable in their parametric range of validity. Comparison with numerical data indicate that 
the cavity result is more accurate over a larger parameter range. 
\end{abstract}

\maketitle

\section{\label{sec:intro}Introduction}

Understanding how competing species coexist in large numbers, as observed in natural ecosystems, is a persistent challenge in theoretical ecology. Several works have addressed the question of 
whether complexity -- for instance measured by biodiversity -- begets stability.~\cite{McA55,Nun80,Mcc00,Hat24,Agu24} 
A closely related question is to understand what conditions favor or even guarantee
the sustainability of large ecological systems with large biodiversity. A modern approach to these questions was pioneered by Robert May, whose Random Matrix Theory (RMT)~\cite{RMT}  approach
predicts that large ecosystems systematically become unstable as their biodiversity increases, unless the variance of interspecies interactions decreases at least inversely proportionally to the number of species.~\cite{May72} 
May's work initiated a new, RMT approach to theoretical ecology.~\cite{May72,Rie89,Opp92,Jan03,All12,All15,Fyo16,Gri17,Fri17,Bun17,Bir18,Bar18,Gal18,Sto18,Ros23,Cle23,Mam24,Jac25,Akj24,Jac26}
Among other, RMT emphasized  a transition between a regime with a single, globally stable equilibrium to 
a phase with multiple, coexisting equilibria as the interspecies interaction strength increases.~\cite{Rie89,Bir18,Ros23} The transition is sensitive to 
how species interact,~\cite{All12,Fri17,Cle23,Ros23} with mutualistic or competitive interactions being generally destabilizing, while predator-prey interactions 
tend to stabilize ecosystems.~\cite{All12,Cle23,Mam24} For reviews of the RMT approach to 
theoretical ecology I advise the reader to read Refs.~\cite{All15,Akj24}

There exist many different mathematical models for population dynamics in ecosystems~\cite{May07} and here I am interested in generalized Lotka-Volterra models defined below in Eq.~\eqref{lv}.
My focus is on the population abundance distribution of its fixed-point solutions. This distribution entails  information on biodiversity, moreover,
asymptotic extinction thresholds can be evaluated from it. 
Fixed-point solutions of Eq.~\eqref{lv} may have negative species abundances.
They do not represent real ecosystem behavior, first, because a negative population has no physical meaning, and second, because these fixed-points
are not reachable by the dynamics of Eq.~\eqref{lv} from the positive hyperoctant, $N_i \ge 0$, $\forall i$. One speaks of {\it unfeasible fixed points}, whose
existence reflects the fact that species  get extinct during the time-evolution of Eq.~\eqref{lv}.
While May focused on fixed-point stability, it has long been recognized that stability and
feasibility must be considered on an equal footing when investigating species coexistence in mathematical models.
~\cite{Gri17,Ros01,Sto16,Bun17,Sto18,Akj22,Mar25,Cle23,Akj24,Jac26,Rob74}
A general consensus has emerged, that feasibility is more restrictive than stability: as the interspecies interaction strength or the number of  species 
increase, species get extinct and fixed-points generically become unfeasible before they become unstable. 

This manuscript follows and complements Ref.~\cite{Jac26} There, I constructed a RMT-based perturbative expansion to calculate the moments of the distribution $P(N_i^*)$ of species abundances for equilibrium fixed-points 
of Eq.~\eqref{lv}. In the regime of weak interaction and large number of species, I found that 
abundances are normally distributed and I calculated  their average $\overline{N}$ and variance $\Sigma_N$. 
These findings extended earlier results~\cite{Gem82,Cle23,Akj24} to a larger parameter space. 
Such a normal distribution has a nonvanishing probability that abundances are negative, $P(N_i^*<0) \ne 0$, and it is tempting to truncate it to positive abundances
only and to consider species with negative abundances as extinct. 
Ref.~\cite{Jac26} showed that this truncation is legitimate as long as 
the number $N_e$ of extinct species is small against
the total number $S$ of species. Truncated Gaussian distributions of abundancies have been found earlier for 
fixed-point solutions of generalized Lotka-Voltera models, via self-consistent numerical methods incorporating feasibility from the start.~\cite{Bun17,Cle23,Barb17} 
The approach of Ref.~\cite{Jac26} establishes  that, for fixed interaction strength,  
fixed points lose feasibility before stability, because its perturbative approach remains within May's regime of stability, where one still finds 
 systematic extinctions when $S \gg 1$ -- i.e. 
even when $\overline{N} \gg \Sigma_N^{1/2}$ and $P(N_i^*<0) \ll 0$, there will be one or more species populating the $N_i^*<0$ tail of the Gaussian distribution,  if
$S$ is sufficiently large.

Here, I first recast the approach of Ref.~\cite{Jac26} in a diagrammatic formalism. I express Feynman rules to calculate $\overline{N}$ and $\Sigma_N$ to any order in the interspecies
interaction strength. Interestingly enough, connections can be made between the perturbation expansion for $\overline N$ with Dyck paths. 
Accordingly, the infinite perturbation expansion for $\overline N$ can be exactly resummed in a closed-form formula via generating functions for Catalan and Narayana numbers. 
I furthermore propose an alternative calculation based on the cavity 
approach previously used in the context of population dynamics in Refs.~\cite{Bun17,Barb17,Bir18,Gal18,Roy19} Unlike these works, however, I do not incorporate
the condition of feasibility in the calculation, which allows me to derive closed-form expressions for both $\overline{N}$ and $\Sigma_N$. The price to pay is that 
the distribution is representative of real ecosystems only as long as it predicts only a small fraction of species with negative abundances. Still, it is fully appropriate to investigate the onset of 
extinctions and the loss of fixed-point feasibility. 

\section{Model and Methodology}\label{sec:model} 

Multispecies Lotka-Volterra models are defined by a the following system of $S$ coupled nonlinear ordinary differential equations~\cite{All15,Akj24}
\begin{equation}\label{lv}
 \dot N_i = N_i \Big( k_i- N_i - \sum_{j=1}^S {\mathbb A}_{ij} N_j \Big) \, , \; i=1, \ldots S \, .
\end{equation}
Eq.~\eqref{lv} generalizes the original, two-species Lotka-Volterra model.~\cite{LV1,LV2} 
It  defines the time-evolution of the abundance $N_i(t)$ of each species $i$, given their growth rate/intrinsic carrying capacity $k_i$ and their interactions with $S-1$ 
coexisting species. For the large ecosystems I consider here, $S \gg 1$ and it is in practice impossible to determine accurately 
the $S^2$ real interaction matrix elements $\mathbb A_{ij}$ determining the growth rate of species $i$ due to its interaction with species $j$.\cite{caveat}
Moreover, even if the $\mathbb A_{ij}$ were all known, determining the dynamics of species abundances, identifying attractors and predicting
species extinctions can only be delegated to numerical
studies already for moderate $S$.~\cite{Gar89,Fus00,Mcg08}

Inspired by works in nuclear,~\cite{Wig58,Dys62,Bro81} atomic~\cite{Fla94} and mesoscopic physics,~\cite{Gor65,Bee97} as well as in 
quantum chaos~\cite{Haake,Jac97,Kos18},
a mathematically powerful alternative has been to assume that the interaction matrix ${\mathbb A}$
belongs to an ensemble of random matrices defined by the statistics 
of its matrix elements.~\cite{RMT,May72,All15,Akj24} Here, I assume that ${\mathbb A}_{ij}$
are normally distributed with 
\begin{subequations}
\label{randommatr}
\begin{eqnarray} 
\langle {\mathbb A}_{ij}\rangle & = & \mu/S \;, \\
\label{xcorr}
\langle {\mathbb A}_{ij} {\mathbb A}_{kl} \rangle & = & \sigma^2 ( \delta_{ik} \delta_{jl}+\gamma \delta_{il} \delta_{jk})/S + (\mu/S)^2 \, .
\end{eqnarray}
\end{subequations}
I furthermore assume that intrinsic carrying capacities are normally distributed with  $ \langle k_i \rangle = \kappa$ and  $\langle k_i k_j \rangle  - \langle k_i \rangle^2 = \chi^2 \delta_{ij}$. The resulting statistical ensemble  is then entirely defined by the five parameters $\mu$, $\sigma$, $\gamma$, $\kappa$ and $\chi$. The first two represent the 
average interespecies interaction and their fluctuations about that. The average introduces an interspecies logistic term in Eq.~\eqref{lv}, accordingly I consider $\mu \ge 0$. 
The cross-diagonal covariance  parameter $\gamma$ quantifies the ratio of predator-prey pairs vs. 
competitive or mutualistic ones:  $\gamma \in [-1,1]$ tunes between exclusively predator-prey behaviors, $\gamma=-1$,
to exclusively symbiotic, i.e. with either competitive or mutualistic behavior, $\gamma=1$.~\cite{caveat1} Finally, $\kappa$ and $\chi$ quantify the average and fluctuations of the intrinsic carrying capacities. 
At the level of Eq.~\eqref{randommatr}, there is no structure in the ecological interaction network and every species interacts one way or another with every other one. 
Ecosystems are characterized by sparse interaction networks~\cite{Pim91,Wil00,Kra03,The10,Del19} and 
 statistical sparsity of ${\mathbb A}$ will be introduced in Section~\ref{sec:sparse}.

I am interested in the statistics of population abundances at fixed-point solutions to Eq.~\eqref{lv}. 
In Ref.\cite{Jac26} I developed a perturbative approach to calculate average, variance and
higher moments of the probability distribution $P(N_i^*)$ of abundances over the RMT ensemble defined by Eq.~\eqref{randommatr}.
The approach is based on a Neumann expansion of the fixed-point equation
\begin{equation}\label{equi2}
{\vec N^*} = (\mathbb{I}+\mathbb{A})^{-1} \vec{k} \, ,
\end{equation}
i.e. I write $\mathbb{A} = \mathbb{M} + \delta \mathbb{A}$ with $\mathbb{M}_{ij}=\mu/S$ containing the average interaction matrix elements and
${\delta \mathbb A}$ containing the fluctuations about that average, i.e. with 
\begin{subequations}
\label{randommatr2}
\begin{eqnarray} \label{rmt2avg}
\langle \delta \mathbb A_{ij} \rangle&=&0 \, \\  
\langle \delta \mathbb A_{ij} \, \delta \mathbb A_{kl}  \rangle &=& \sigma^2 ( \delta_{ik} \delta_{jl}+\gamma \delta_{il} \delta_{jk})/S \, .
\end{eqnarray}
\end{subequations}
Expanding the right-hand side of Eq.~\eqref{equi2} in  $\delta {\mathbb A}$, I obtain
\begin{equation}\label{expandN2}
\vec{N^*} =   \big({\mathbb I}-\frac{{\mathbb M}}{1+\mu}\big) \, \sum_{p=0}^\infty  (-1)^p \,  \Big[ \delta \mathbb{A} \big({\mathbb I}-\frac{{\mathbb M}}{1+\mu}\big)\Big]^p \, \vec{k}  \, ,
\end{equation}
where I used $(\mathbb{I+M})^{-1}=\mathbb{I}-\mathbb{M}/(1+\mu)$. Convergence of the Neumann expansion requires that the spectral radius $\rho=\sigma (1+|\gamma|)$
of $\delta \mathbb{A} \big({\mathbb I}-\frac{{\mathbb M}}{1+\mu}\big)$ is smaller than one. Accordingly, Eq.~\eqref{expandN2} is valid as long as 
$\sigma < (1+|\gamma|)^{-1}$, which is May's stability border~\cite{May72} including the elliptic complex eigenvalue distribution with finite $\gamma$.~\cite{Som88}
Using a Neumann expansion to extract species abundances has been done before in Ref.~\cite{Pet20} under 
an assumption of statistical independence. A geometric
resummation was obtained for $\gamma=0$, $\mu=0$ and $\chi=0$. 

My strategy here is to use Eq.~\eqref{randommatr2} to calculate the average and variance 
of the abundance distribution. For more details, the
reader is referred to Ref.\cite{Jac26} Earlier, similar approaches,~\cite{Bun17,Barb17,Roy19,Cle23,Akj24}  include constraints on fixed-point 
feasibility, i.e. $N_i^* \ge 0$, however the present approach does not.
I calculate
moments of the unconstrained distribution $P(N_i^*)$, which may allow for $N^*_i <0$ for some species, in some parameter range. 
The advantage is that one ends up with closed-form expressions that can be analytically evaluated, instead
of self-consistent equations requiring a numerical solution.~\cite{Bun17,Barb17,Roy19} The price to pay is that 
the emergence of extinctions is inferred {\it a posteriori} from the calculated moments, giving the weight of negative values $P(N_i^*<0)$ of the resulting Gaussian distribution.
In Ref.\cite{Jac26} I showed that the fixed-point solution of Eq.~\eqref{equi2} -- with possibly few negative abundances -- approximates the real fixed-point of the dynamics 
with an accuracy ${\cal O}(N_e/S)$ that is excellent
as long as $N_e \ll S$. 

Eq.~\eqref{expandN2} gives an expansion of $N_i^*$ in powers of $\sigma$ and $1/(1+\mu)$. 
Ref.\cite{Jac26} calculated the fixed-point abundance average $\overline N$ 
up to and including all terms of order $\sigma^4$ with all $1/(1+\mu)$-dependences accounted for, as well as 
 the term of order $\sigma^6$ for $\mu=0$. While the method could in principle express all $\mu=0$ terms up to any order in $\sigma$, the combinatorial prefactor
for each term was not explicitly given there. Furthermore, it was unclear how to calculate all $1/(1+\mu)$-dependences. In Section~\ref{sec:pert}, I give the exact form for all combinatorial prefactors and include all $1/(1+\mu)$-terms at each order in $\sigma$. I further resum all contributions to obtain a closed-form 
expression for $\overline N$. In Section~\ref{sec:cav}, I next discuss a second approach 
which uses the cavity method of Refs.~\cite{Bun17,Barb17,Bir18,Bar18,Gal18,Roy19} albeit without the constraint of feasibility. 
The approach also delivers a closed-form expression.
From the Gaussian distribution of species abundances, Ref.\cite{Jac26} conjectured a single-parameter scaling
law for the emergence of extinctions, and  this single parameter can be written exactly in a particularly compact form using the cavity method.
This is discussed in Section~\ref{sec:scaling}.
The resummed, perturbative expression of Section~\ref{sec:pert}  is parametrically
different from the cavity expression, however the two differ by at most few percents in the regime of validity of the theory and are practically undistinguishable as long as $N_e/S \ll 1$. 
A direct comparison between the two approaches is briefly discussed in Section~\ref{sec:compare}. Above, I mentioned that the full RMT assumption made in Eq.~\eqref{randommatr} does not capture 
the structure of 
ecological networks. I show In Section~\ref{sec:sparse} how both perturbative and cavity approaches can be extended to account for sparsity in the interaction matrix, accounting for species that coexist without interacting. 
Section~\ref{sec:nums} gives numerical illustrations to those analytical results. Between Ref.\cite{Jac26} and the present paper, I have interacted with several 
generative AI chatbots. I describe these interactions in Section~\ref{sec:ai}, and the corresponding chats are public and can be consulted.~\cite{claude,gemini,chatgpt}
Conclusions and possible extensions of this work are briefly discussed in Section~\ref{conclusion}.

\section{Perturbative Approach}\label{sec:pert}
\subsection{Average Abundance}\label{sec:pert_avg}

Taking the RMT average of Eq.~\eqref{expandN2} requires to calculate the following terms,
\begin{eqnarray}\label{pairingp}
\left\langle \Big[ \delta \mathbb{A} \big({\mathbb I}-\frac{{\mathbb M}}{1+\mu}\big)\Big]^p_{j_0j} \, k_j \right \rangle = \sum_{j's} 
\left\langle \delta \mathbb{A}_{j_0j_1} \big({\mathbb I}-\frac{{\mathbb M}}{1+\mu}\big)_{j_1j_2} \delta \mathbb{A}_{j_2j_3} \big({\mathbb I}-\frac{{\mathbb M}}{1+\mu}\big)_{j_3j_4}  \ldots 
\delta \mathbb{A}_{j_{2p-2}j_{2p-1}} \big({\mathbb I}-\frac{{\mathbb M}}{1+\mu}\big)_{j_{2p-1} j} \right \rangle \, \langle k_j \rangle \, , \qquad
\end{eqnarray}
where I assumed that the matrix elements of $\delta \mathbb{A}$ are uncorrelated with the intrinsic carrying capacities $k_i$.
Two things need to be taken care of. First, RMT averages need to be taken over the matrices $\delta \mathbb{A}$. 
Second, contributions from 
$\mu-$dependent matrices ${\mathbb I}-{\mathbb M}/(1+\mu)$ need to be calculated. Each such term gives either of two contributions
\begin{eqnarray}\label{1mu}
{\mathbb I}-{\mathbb M}/(1+\mu) \rightarrow \left \{
\begin{array}{cc}
1 + {\cal O}(1/S) \, , &  \; {\rm or} \\
1/(1+\mu) \, , &   
\end{array}
\right.
\end{eqnarray}
depending on constraints on index summation introduced by the RMT contractions of the $\delta {\mathbb A}$ matrices.

\begin{figure}[t!]
\center
\includegraphics[width=0.4 \columnwidth]{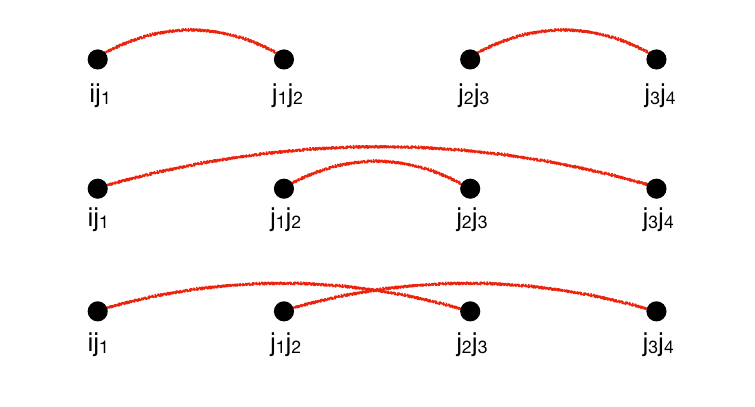}
\caption{Top and middle: the $C_2=2$ noncrossing contractions that contribute to $\langle \delta {\mathbb A}_{ij_1} \delta {\mathbb A}_{j_1j_2} \delta {\mathbb A}_{j_2j_3} \delta {\mathbb A}_{j_3j_4} \rangle$.
Each fluctuation matrix is represented by a black dot, with the corresponding two indices indicated. Pair-contractions are represented by the blurred red lines. 
The bottom contraction corresponds to the third term on the right-hand side of Eq.~\eqref{pairing4}. It gives a subdominant contribution ${\cal O}(1/S)$. 
This is so because it requires to equate pairs of indices at least three times, leaving only a single index
summation in Eq.~\eqref{pairing4}.
\label{fig1}}
\end{figure}

I first sketch the calculation of the average in Eq.~\eqref{pairingp} for $\mu=0$. Performing the RMT average is tantamount to summing over all 
pair-contractions of fluctuation matrices $\delta \mathbb{A}$. For each $p$-term, there are $p$ such matrices and the RMT average, Eq.~\eqref{randommatr2}, selects out terms with even 
powers, $p=2 m$ with $m \in \mathbb{N}$.
Furthermore, the thermodynamic limit $S \rightarrow \infty$ isolates
non-crossing pair contractions over indices $j_0, j_1, \ldots$ in Eq.~\eqref{pairingp}, because crossing contractions lead to factors ${\cal O}(1/S)$. 
The simplest way to see this is to consider the $p=4$ term, 
where there are three possible contractions. They are depicted in Fig.~\ref{fig1}, where black dots labelled with indices $ij$ denote the matrix element 
$\delta \mathbb{A}_{ij}$
and red lines indicate pair-contractions. The three contractions give 
\begin{eqnarray}\label{pairing4}
\sum_{j_1,j_2,j_3,j_4}
\langle \delta {\mathbb A}_{ij_1} \delta {\mathbb A}_{j_1j_2} \delta {\mathbb A}_{j_2j_3} \delta {\mathbb A}_{j_3j_4} \rangle &=&
(\sigma^4/S^2) \sum_{j_1,j_2,j_3,j_4}[ \gamma^2 (\delta_{ij_2} \delta_{j_2j_4} + \delta_{j_1j_3} \delta_{ij_4}) +\delta_{ij_2} \delta_{j_1j_3}\delta_{j_2j_4}] 
= \sigma^4 (2 \gamma^2+ S^{-1})\, .
\end{eqnarray}
The crossing, third contraction is subdominant in $1/S$, because it gives rise to three Kronecker-$\delta$, leaving only one index summation in Eq.~\eqref{pairing4}, whereas two are needed to cancel the $S^2$ denominator in Eq.~\eqref{pairing4}. I obtain  two rules that, first, crossing contractions (as in the bottom panel of Fig.~\ref{fig1}) give no contribution in the thermodynamic limit and second, 
contractions of pairs of consecutive $\delta \mathbb{A}$ give a prefactor $\gamma \sigma^2/S$ and remove one index summation on the right-hand side of 
Eq.~\eqref{pairingp}. Contributions that survive the thermodynamic limit for the $p=2m$-term have therefore $p/2=m$ non-crossing contractions that leave exactly 
$p/2=m$ index summations. 

Contributing pair-contractions in the thermodynamic limit can therefore be visualized as planar graphs, i.e. with no crossing. 
The combinatorial factor determining the number of such planar graphs for $p=2 m$ is given by the Catalan number $C_m = (2m)!/[(m+1)!m!]$.~\cite{catalan}
Therefore, for $\mu=0$, the average abundance is given by the generating function $C(x)$ of the Catalan numbers,
\begin{equation}\label{expandN3}
\overline{N}_{\mu=0} = \kappa \, \sum_{m=0}^\infty  C_m \; (\gamma \sigma^2)^m = \kappa \, \frac{1-\sqrt{1-4 \gamma\sigma^2}}{2 \gamma \sigma^2} = C(\gamma \sigma^2) \,  \, .
\end{equation}
Eq.~\eqref{expandN3} generalizes Eq.({\color{red}4}) in Ref.\cite{Jac26} to all orders in $\sigma$ for $\mu=0$.
Each $p$-term in Eq.~\eqref{expandN2} with $\mu=0$ is recovered by Taylor-expanding the squareroot in Eq.~\eqref{expandN3}, which explicitly determines its
range of validity, $|\gamma \sigma^2| < 1/4$. However, for the full range $\gamma \in [-1;1]$, this condition is less restrictive than the convergence radius of the 
Neumann expansion in Eq.~\eqref{expandN2}, $\sigma < (1+|\gamma|)^{-1}$, which therefore determines the range of validity of Eq.~\eqref{expandN3}. 

\begin{figure}[h]
\center
\includegraphics[width=0.45 \columnwidth]{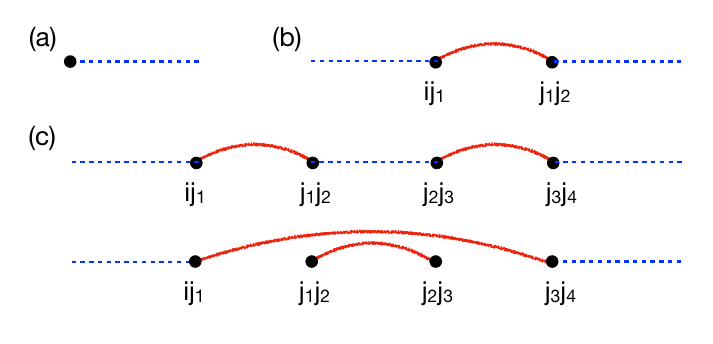}
\caption{(a) Dressed $\delta {\mathbb A}$-vertex (black dot), including the postfactor ${\mathbb I}-{\mathbb M}/(1+\mu)$ in Eq.~\eqref{expandN2} (dashed blue line).
(b) The sole contribution to the $p=2$ term in Eq.~\eqref{expandN2} at finite $\mu \ne 0$. Pairing between the two vertices effectively removes a dashed blue line,
$[\mathbb{I} - {\mathbb M}/(1+\mu)]_{j_1j_1} \rightarrow 1$. (c) The two contributions to the $p=4$ term in Eq.~\eqref{expandN2} at finite $\mu \ne 0$. In (b) and (c), the dashed blue lines to the left of the leftmost vertex arise from the first
prefactor $\mathbb{I} - {\mathbb M}/(1+\mu)$ on the right-hand side of Eq.~\eqref{expandN2}.
\label{fig2}}
\end{figure}

\begin{figure}[h!]
\center
\includegraphics[width=0.6 \columnwidth]{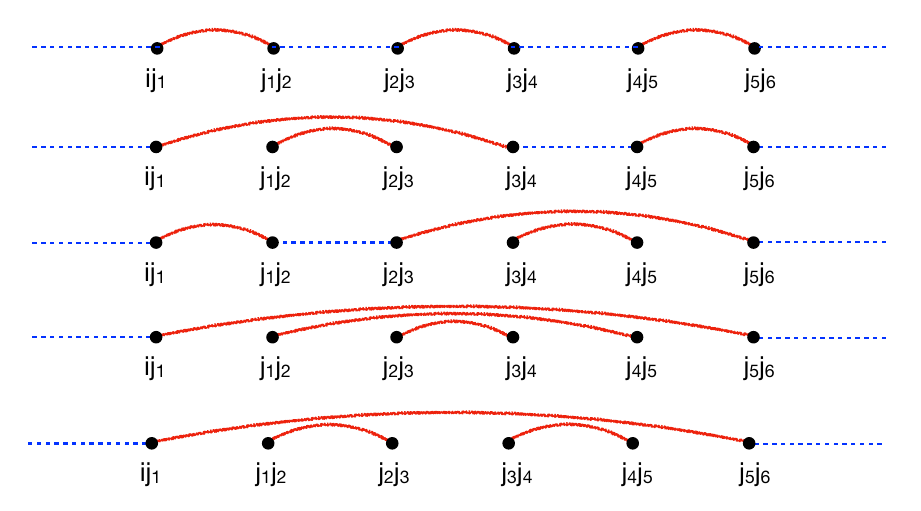}
\caption{The $C_3=5$ dominant pair-contractions that contribute to the $p=6$ term in Eq.~\eqref{expandN2} in the thermodynamic limit. The blue dashed lines represent terms which do not suffer from constrained index summation and
which therefore contribute a factor ${\mathbb I}-{\mathbb M}/(1+\mu) \rightarrow 
1/(1+\mu)$ because they lie between or outside of disconnected contraction blocks.
\label{fig3}}
\end{figure}

I next consider finite values $\mu \ge 0$. Once indexed, factors to be additionally included are represented as the dashed blue line in panel (a) of Fig.~\ref{fig2}.
They read $[{\mathbb I}-{\mathbb M}/(1+\mu)]_{ij} = \delta_{ij} - (\mu/S)/(1+\mu)$. The second term  is of higher 
order in $S^{-1}$ than the first one, however, it is not index-restricted. Therefore, $1/(1+\mu)$-factors emerge when there is an unrestricted summation over $i$, $j$ or both,
$\sum_i [{\mathbb I}-{\mathbb M}/(1+\mu)]_{ij} = \sum_i [\delta_{ij} - (\mu/S)/(1+\mu)] = 1-\mu/(1+\mu) = 1/(1+\mu)$.
Key to understand quantitatively how factors $1/(1+\mu)$ emerge is to realize that contracted planar graphs 
define blocks of connected components. For instance, the top contraction in Fig.~\ref{fig1} defines two disconnected blocks while the middle contraction defines one block.
Inside a block, index summations are restricted by pair-contractions, which effectively removes the $\mu$-dependence ${\mathbb I}-{\mathbb M}/(1+\mu) \rightarrow 1$.
This is most easily seen for the $p=2$ contribution represented in panel (b) in Fig.~\ref{fig2},
\begin{eqnarray}
\overline N_{i;p=2}
&=&  \kappa \, \sum_{j's} \Big\langle  \delta \mathbb{A}_{j_1j_2} \delta \mathbb{A}_{j_3j_4} \Big\rangle \; \big({\mathbb I}-\frac{{\mathbb M}}{1+\mu}\big)_{ij_1} \big({\mathbb I}-\frac{{\mathbb M}}{1+\mu}\big)_{j_2j_3}   \big({\mathbb I}-\frac{{\mathbb M}}{1+\mu}\big)_{j_4j} \nonumber \\
&=&  \kappa\, \frac{\gamma \sigma^2}{S} \sum_{j's} \delta_{j_1j_4} \delta_{j_2j_3} \, [\delta_{ij_1}-(\mu/S)/(1+\mu)] \; [\delta_{j_2j_3}-(\mu/S)/(1+\mu)] \;  [\delta_{j_4j}-(\mu/S)/(1+\mu)] \nonumber \\
&=& \frac{\kappa \gamma \sigma^2}{(1+\mu)^2} \; .
\end{eqnarray}
The dominant pairing $j_1=j_4$, $j_2=j_3$ 
suppresses the $\mu$-dependence of the ${\mathbb I}-{\mathbb M}/(1+\mu)$-term between the two paired $\delta {\mathbb A}$ vertices, $\delta_{j_2j_3} [ \delta_{j_2j_3}  - (\mu/S)/(1+\mu) ]\rightarrow \delta_{j_2j_3} + {\cal O}(1/S)$.
This generalizes to all blocks of any size, inside which $\mu$-dependences disappear. This is shown in panel (c) in Fig.~\ref{fig2} for the $p=4$ term and 
in Fig.~\ref{fig3} for the $p=6$ term, whose contribution is given by $C_{3}=5$ terms. 

Putting all this together, we have that pair-contractions diagrammatically determine the number of disconnected blocks, which in their turn determine the number of 
factors $(1+\mu)^{-1}$. Instead of Eq.~\eqref{expandN3} I write
\begin{equation}\label{expandN4}
\overline N = \frac{\kappa}{1+\mu} \, \sum_{m=0}^\infty  \sum_{k=0}^m {\cal N}(m,k) (\gamma \sigma^2)^m (1+\mu)^{-k} \, , 
\end{equation}
where ${\cal N}(m,k)$ gives the number of diagrams with $p=2m$ paired vertices
forming $k$ blocks. Such diagrams can be encoded as so-called Dyck paths of semilength $m$ with $k$ peaks, the number of which is given by the 
Narayana numbers ${\cal N}(0,0)=1$, ${\cal N}(m,0)=0$ and
${\cal N}(m,k) = m^{-1} \binom{m}{k} \binom{m}{k-1}$ otherwise.~\cite{catalan,Deu99} As a consistency check, note that for $\mu=0$, equating Eq.~\eqref{expandN3}
and Eq.~\eqref{expandN4} requires $\sum_{k=0}^m {\cal N}(m,k) = C_m$, which is a well-known relation between the Narayana and Catalan numbers.~\cite{catalan}
The double sum on the right-hand side of Eq.~\eqref{expandN4} can then be resummed with the help of the generating function $S(x,y)$ of the Narayana numbers,
\begin{equation}
S(x,y) = \sum_{m,k \ge 0} {\cal N}(m,k) x^m y^k = \frac{1+x-xy-\sqrt{1-2x+x^2-2xy-2x^2y+x^2y^2}}{2 x} \, .
\end{equation}
Identifying $x=\gamma \sigma^2$ and $y=(1+\mu)^{-1}$, I obtain the generalization of Eq.~\eqref{expandN3} to nonvanishing values of $\mu$,
\begin{equation}\label{expandN5}
\overline N = \frac{\kappa}{1+\mu} \, S(\gamma \sigma^2,(1+\mu)^{-1}) = \frac{\kappa}{1+\mu} \,   \frac{1+\gamma \sigma^2 \mu/(1+\mu) - \sqrt{1-2 \gamma\sigma^2 (2+\mu)/(1+\mu) + (\gamma \sigma^2)^2 \mu^2/(1+\mu)^2}}{2 \gamma \sigma^2} \,  \, .
\end{equation}
This is a remarkable result, as it exactly resums an infinite perturbation expansion in an exact closed-form formula. 
The range of validity of Eq.~\eqref{expandN5} remains $\sigma < 1/(1+|\gamma|)$, even for $\mu >0$, because $\mu$ does not change the spectral radius of 
$\delta {\mathbb A} \, [{\mathbb I}-{\mathbb M}/(1+\mu)]$.

I conclude this paragraph by writing down the Feynman rules for calculating the contribution to order $\sigma^p$ to $\overline N$: \\

(1) only even contributions survive the RMT average, thus $p=2 m$;

(2) diagrams are made of $p$ vertices -- they are represented by the black dots in Figs.~\ref{fig1}, \ref{fig2} and \ref{fig3};

(3) each non-crossing pairing of vertices defines one diagram to be considered -- pairings are indicated by the blurred red lines in Figs.~\ref{fig1}, \ref{fig2} and \ref{fig3};

(4) pairings define blocks in non-crossing planar diagrams and factors of $1/(1+\mu)$ dress the diagrams to their left and right, as well as in between each block
-- these are the dashed blue lines in Figs.~\ref{fig2} and \ref{fig3};

(6) each blurred red line contributes a factor $\gamma \sigma^2$;

(7) each dashed blue line contributes a factor $1/(1+\mu)$;

(8) the resulting contribution is finally mutliplied by $\kappa$.\\

\subsection{Variance of the Abundance Distribution}\label{var_pert}

The second moment of the abundance distribution is given by 
\begin{eqnarray}\label{var1}
\langle (N_i^*)^2 \rangle &=& 
\sum_{j,l} \langle [(\mathbb{I}+\mathbb{A})^{-1}]_{ij} [(\mathbb{I}+\mathbb{A})^{-1}]_{il} \rangle \, \langle k_j k_l \rangle \, .
\end{eqnarray}
I sketch how contributions to each order in $\sigma$ can be calculated 
from easily constructed sets of diagrams. Unlike for the average abundance, I am able to resum the infinite series for $\Sigma_N$ 
and below, I truncate the perturbative expansion at ${\cal O}(\sigma^6)$, with each term including its full $\mu-$dependence.

As for the average abundance, I use
\begin{eqnarray}
 [(\mathbb{I}+\mathbb{A})^{-1}]_{ij}  &=& \sum_{j_1,j} 
\big({\mathbb I}-\frac{{\mathbb M}}{1+\mu}\big)_{ij_1} \, \sum_{p=0}^\infty  (-1)^{p} \,  
\Big[ \delta \mathbb{A} \big({\mathbb I}-\frac{{\mathbb M}}{1+\mu}\big)\Big]^{p}_{j1j}  \, ,
\end{eqnarray}
to rewrite Eq.~\eqref{var1} as
\begin{eqnarray}\label{var2}
\langle (N_i^*)^2 \rangle &=&  \sum_{j_1,j} \, \sum_{l1,l}
\big({\mathbb I}-\frac{{\mathbb M}}{1+\mu}\big)_{ij_1} \, 
\big({\mathbb I}-\frac{{\mathbb M}}{1+\mu}\big)_{il_1} \, \sum_{p_1,p_2=0}^\infty  (-1)^{p_1+p_2} \,  \Big\langle
\Big[ \delta \mathbb{A} \big({\mathbb I}-\frac{{\mathbb M}}{1+\mu}\big)\Big]^{p_1}_{j_1j} 
\Big[ \delta \mathbb{A} \big({\mathbb I}-\frac{{\mathbb M}}{1+\mu}\big)\Big]^{p_1}_{l_1l} \Big \rangle \, \langle k_j k_l \rangle \, . \qquad
\end{eqnarray}
Again assuming that the distribution of interaction matrices $\delta {\mathbb A}$ is not
correlated with the distribution of intrinsic carrying capacities $k_i$, I obtain a factor
$\langle k_j k_l \rangle = \kappa^2 + \delta_{jl} \chi^2$.

\begin{figure}[t]
\center
\includegraphics[width=0.8 \columnwidth]{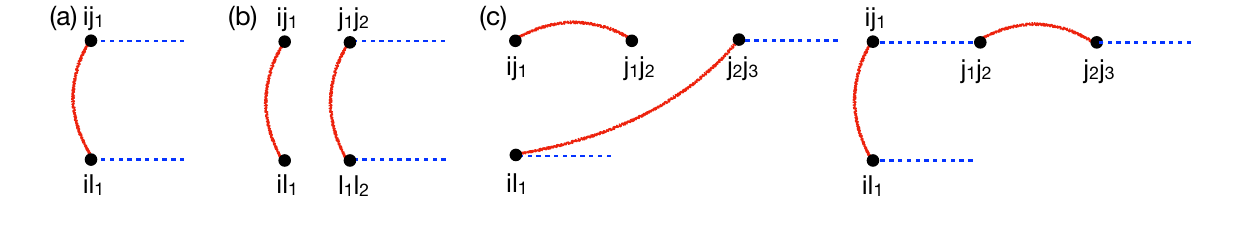}
\caption{Diagrams contributing to the variance $\Sigma_N(\chi=0)$ of species abundances: (a) to order $\sigma^2$, with $p_1=p_2=1$ in Eq.~\eqref{var2}, and
to order $\sigma^4$ (b) with $p_1=p_2=2$ and (c) with $p_1=3$ and $p_2=1$. According to the rules listed in the main text, these diagrams contribute 
(a) $\sigma^2/(1+\mu)^2$, (b) $\sigma^4/(1+\mu)^2$, (c) $\gamma \sigma^4[(1+\mu)^{-2} + (1+\mu)^{-3}]$. 
\label{fig4}  }
\end{figure}

Fig.~\ref{fig4} shows the diagrams contributing to $\Sigma_N$, 
up to and including order 
$\sigma^4$. I use them to illustrate the formalism and write down the rules for the perturbative calculation of $\Sigma_N$.
First, from Eq.~\eqref{randommatr2},  the terms in Eq.~\eqref{var2} that survive RMT average are those with even sum $p_1+p_2$. 
Second, Eq.~\eqref{var2} defines  two separate sets of vertices, one with $p_1$, one with $p_2$ vertices. 
Diagrammatically, I represent each set as a horizontal line starting at $i$ to the left. Fig.~\ref{fig4} (a) illustrates the situation for 
$p_1=p_2=1$ and Fig.~\ref{fig4} (b) for $p_1=3$ and $p_2=1$. Contributions to the variance correspond 
to connecting contractions, i.e. where at least one index pairing connects the two sets of vertices. Third, each pairing within a set gives a contribution $\gamma \sigma^2/S$ 
while connecting pairings, i.e. those connecting two vertices on different subsets, give a contribution 
$\sigma^2/S$. Fourth, as for the abundance average, contractions define blocks inside which no $\mu$-dependent factor survives. Fifth, $\mu$-dependent contributions
depicted by the dashed blue lines in Fig.~\ref{fig4} exist between blocks that have no connecting index pairing to their right. 
This follows 
from constraints on index summation induced by connecting contractions, which lead to terms $\propto {\rm Tr}\left[[{\mathbb I}-{\mathbb M}/(1+\mu)]^q\right] = S [1+{\cal O}(S^{-1})]$, $q \ge 1$,
with no $\mu$-dependent terms to leading order in $S$. Sixth, leftmost blue lines corresponding to the prefactor 
$\big({\mathbb I}-\frac{{\mathbb M}}{1+\mu}\big)_{ij_1} \, \big({\mathbb I}-\frac{{\mathbb M}}{1+\mu}\big)_{il_1}$ in Eq.~\eqref{var2} give no $\mu-$dependence. This results from 
them sharing the index $i$, together with connecting contractions.

From the diagrams in Fig.~\ref{fig4} and the 21 diagrams contributing at order $\sigma^6$ (not shown)
I obtain
\begin{equation}\label{sigma1}
\Sigma_N(\chi \equiv 0)=  \frac{\kappa^2}{(1+\mu)^2} \left[ \sigma^2+ \sigma^4 [1+2 \gamma +\frac{2 \gamma}{1+\mu} ] 
+\sigma^6 [1 + 4 \gamma + 5 \gamma^2 +\frac{2 \gamma + 6 \gamma^2}{1+\mu}  + \frac{3 \gamma^2}{(1+\mu)^2} ]
\right] \, . 
\end{equation}

This expression gives the complete $\mu$-dependence for each term. Eq.~({\color{red} 5}) in Ref.~\cite{Jac26} instead gave a truncated expansion in $\mu$ and $\sigma^2$,
including terms $\mu^a \sigma^{2 b}$, $a+2b \le 6$. Moreover, the six rules given above that identify diagrammatic contributions make it in principle possible 
to calculate perturbative corrections to any order in $\sigma$. These are significant 
improvements from Ref.~\cite{Jac26} however the blow-up of the number of diagrams to be considered at higher orders makes it computationally heavy to calculate the order $\sigma^8$ already.
I therefore stopped the expansion at order $\sigma^6$, for which 21 diagrams contribute. There are more than 100 of them at order $\sigma^8$.
I have not been able to derive combinatorial factors for each contributions as was done for $\overline N$ in Section~\ref{sec:pert_avg}. Therefore the Feynman rules for calculating 
perturbative contributions are more important for $\Sigma_N$ than for $\overline N$. For the $\kappa-$dependent contributions to the variance, they read \\

(1) only even contributions survive the RMT average, thus $p_1+p_2=2 m$ in Eq.~\eqref{var2};

(2) diagrams are made of two parallel lines of vertices, one with $p_1$ the other with $p_2$ vertices -- see Fig.~\ref{fig4};

(3) non-crossing pairings of vertices to be considered must connect the two parallel lines with at least one pairing; they each define a diagrammatic contributions to $\Sigma_N$;  

(4) factors of $1/(1+\mu)$ dress the diagrams on both lines of vertices to their right, as well as in between pairing blocks, as long as the latter lie to the right of the crossing(s)
connecting the two lines of vertices;

(5) each blurred red line contributes a factor $\gamma \sigma^2$ if it pairs within a line and $\sigma^2$ if it connects vertices on different lines;

(6) each dashed blue line contributes a factor $1/(1+\mu)$;

(7) the resulting contribution is mutliplied by $\kappa^2$;\\

More diagrams contribute to $\Sigma_N$ when $\chi \ne 0$. Identifying and calculating them is greatly facilitated by the index contraction $\delta_{jl}$ arising from $\langle k_j k_l \rangle$ in Eq.~\eqref{var2}.
Because the initial ($i$) and the final ($j=l$) species index are fixed and equal in the two lines of vertices, this induces summation constraints that cancel all $\mu-$dependencies. 
One is therefore left with evaluating the following expression
\begin{eqnarray}\label{var3}
\langle (N_i^*)^2 \rangle_\chi &=&  \chi^2 \sum_{j} \, \sum_{p_1,p_2=0}^\infty  (-1)^{p_1+p_2} \,  \Big\langle
\Big[ \delta \mathbb{A} \Big]^{p_1}_{ij} 
\Big[ \delta \mathbb{A} \Big]^{p_1}_{ij} \Big \rangle  \, . \qquad
\end{eqnarray}

\begin{figure}[h!]
\center
\includegraphics[width=0.5 \columnwidth]{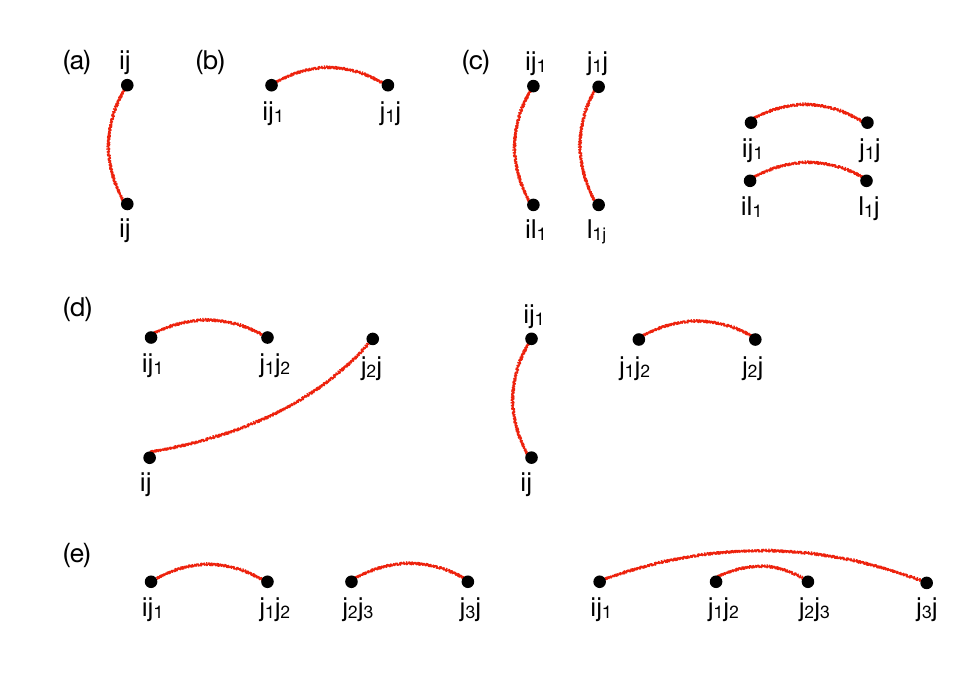}
\caption{Additional, $\chi-$dependent diagrams contributing to the variance $\Sigma_N(\chi \ne0)$ of species abundances: to order $\sigma^2$ (a) with $p_1=p_2=1$ in Eq.~\eqref{var3}
and (b) with $p_1=2$, $p_2=0$; and
to order $\sigma^4$ (c) with $p_1=p_2=2$, (d) with $p_1=3$ and $p_2=1$ and (e) with $p_1=4$ and $p_2=0$. Contributions with $p_1 \ne p_2$ are counted with multiplicity 2.
According to the rules listed in the main text, these diagrams contribute (including their multiplicity)
(a) $\sigma^2$, (b) $2 \gamma \sigma^2$, (c) $\sigma^4 (1+\gamma^2)$, (d) $4 \gamma \sigma^4$, and (e) $2 \gamma \sigma^4$.
\label{fig5}  }
\end{figure}

Feynman rules for the corresponding, $\chi-$dependent diagrams read \\

(1) only even contributions survive the RMT average, thus $p=p_1+p_2=2 m$;

(2) diagrams are made of two parallel lines of vertices, one with $p_1$ the other with $p_2$ vertices -- see Fig.~\ref{fig4};

(3) non-crossing pairings of vertices have to be considered, however they do not need to connect the two parallel lines; they each define a diagrammatic contributions to $\Sigma_N$;  

(4) each blurred red line contributes a factor $\gamma \sigma^2$ if it pairs within a line and $\sigma^2$ if it connects vertices on different lines;

(5) the resulting contribution to the variance is mutliplied by $\chi^2$.\\

Diagrams up to and including the order $\sigma^4$ are shown in Fig.~\eqref{fig5}. Including multiplicity, there are 35 diagrams to be considered at order $\sigma^6$  and I do not show
them here. I obtain the variance
\begin{eqnarray}\label{sigma2}
\Sigma_N(\chi \ne 0)  &=&  \frac{\kappa^2}{(1+\mu)^2} \left[ \sigma^2+ \sigma^4 [1+2 \gamma +\frac{2 \gamma}{1+\mu} ] 
+\sigma^6 [1 + 4 \gamma + 5 \gamma^2 +\frac{2 \gamma + 6 \gamma^2}{1+\mu}  + \frac{3 \gamma^2}{(1+\mu)^2} ] 
\right] \nonumber \\
&&+ \chi^2 \left[ 1+(1+2 \gamma) \sigma^2 +(1+4 \gamma + 5 \gamma^2) \sigma^4 +(1+6 \gamma + 14 \gamma^2 + 14 \gamma^3) \sigma^6 \right] \, .
\end{eqnarray}
This expression extends Ref.~\cite{Jac26} as it resums all orders in $\mu$ for each calculated term. The $\chi$-dependent terms have no $\mu$-dependence and are the same as already 
obtained in Ref.~\cite{Jac26}

\section{Cavity Method}\label{sec:cav}

In Section~\ref{sec:pert_avg}, the perturbative approach or Ref.~\cite{Jac26} has been extended to deliver an exact closed-form expression resumming all order contributions to $\overline N$,
Eq.~\eqref{expandN5}. I have been unable, however, to run a similar program for $\Sigma_N$.
In this section, I use the cavity method to calculate both $\overline N$ and $\Sigma_N$. The method delivers closed-form expressions for both quantities. 

\subsection{Average Abundance}\label{sec:average_cav}

The cavity method is a mean-field approximation. It has been introduced in the context of spin glasses and the long-range interacting Sherrington-Kirkpatrick model.~\cite{Mez86} It is based on a linear response approximation to the removal of a single spin site -- in our case, of a single species. The associated susceptibilities
are evaluated within the double approximation that (i) only the diagonal susceptibilities matter and (ii) they are identical. One then obtains equations
relating the average local magnetization -- or in our case the average species abundance -- to the susceptibility.
The method has been imported to theoretical ecology first by Rieger~\cite{Rie89} and Opper and Diederich.~\cite{Opp92} It has more recently been applied to large, multispecies ecosystems modelled by 
generalized Lotka-Volterra or similar models in Refs.~\cite{Bun17,Akj24,Barb17,Roy19} To make a long story short, one  derives
mean-field, self-consistent equations for surviving species abundances that are solved numerically. The difficulty arises from the immediate, {\it a priori}  
inclusion of 
extinctions in the calculation as constraints $N_i^*=0$ for each extinct species.~\cite{Bun17,Akj24,Barb17,Roy19} They need to be self-consistently determined for a given set of model parameters.
Following Ref.\cite{Jac26} I circumvent this difficulty by focusing on the statistics of abundances for fixed points of Eq.~\eqref{lv}, i.e. full solutions of Eq.~\eqref{equi2}
with possibly species with unphysical negative abundances. Only once average and variance of the species distribution are calculated do I evaluate the 
number $N_e$ of extinct species as those lying in the negative sector of the abundance distribution. The method is consistent as long as $N_e \ll S$. As a matter of fact, I showed in Ref.\cite{Jac26} that the surviving abundances for
unfeasible fixed-points  [i.e. solutions to Eq.~\eqref{equi2} with some $N_i^* < 0$] and for the physical ones [i.e. those obtained by evolving 
Eq.~\eqref{lv} until $N_i^* \rightarrow 0$ for some species] differ only by negligible amounts, ${\cal O}(N_e/S)$, as long as $N_e/S \ll 1$. Compared to Refs.~\cite{Bun17,Barb17,Akj24,Roy19} the present approach has the advantage to deliver closed-form expressions for the abundance 
average and variance. The price to pay is that, to investigate extinctions, it is valid only in the weak interaction regime, as long as $N_e/S \ll1$. 

The cavity method compares equilibrium abundances for the full system, i.e. solutions to 
\begin{equation}\label{eq1}
N_i^* = k_i - \sum_l {\mathbb A}_{il} N_l^* \, , 
\end{equation}
to those obtained after the removal of one, say the $j^{\rm th}$ species, 
\begin{equation}
N_i^{(j)}= k_i - \sum_{l \ne j} {\mathbb A}_{il} N_l^{(j)} \, .
\end{equation}
The abundance shifts $\delta N_i = N_i^* - N_i^{(j)}$, $i \ne j$ read
\begin{equation}\label{cavity}
\delta N_i = -N_j^* \sum_{l \ne j} \varphi_{il} {\mathbb A}_{lj} \, , 
\end{equation}
where I introduced the susceptibilities $\varphi_{il} = \partial N_i^{(j)}/\partial k_l=(1+{\mathbb A}^{(j)})^{-1}_{il}$, with ${\mathbb A}^{(j)}$ obtained from
${\mathbb A}$ by removing the $j^{\rm th}$ line and row. 
The cavity approximation first neglects off-diagonal susceptibilities, $\varphi_{il} =0,$ $i \ne l$. This is qualitatively justified because the corresponding terms in 
Eq.~\eqref{cavity} sum over the zero-average random variable ${\mathbb A}_{lj}$ of typical magnitude ${\cal O}(1/\sqrt{S})$. Under the assumption 
that ${\mathbb A}_{lj}$ and $\varphi_{il}$ 
are uncorrelated, the resulting sum is negligible in the thermodynamic limit. The second approximation it a standard
mean-field assumption that the surviving susceptibilities $\varphi_{ii} = \varphi$, $\forall i$. I obtain
\begin{equation}\label{cavity2}
\delta N_i = -N_j^* \varphi {\mathbb A}_{ij} \;, \;\;\; {\rm i.e.} \;\;\;\;\;\;\;\;\; N_i^* = N_i^{(j)} -N_j^* \varphi {\mathbb A}_{ij} \, .
\end{equation}
Substituting this latter equation into 
Eq.~\eqref{eq1} leads to
\begin{equation}\label{cav0}
N_i^* = k_i - \sum_{l \ne j} {\mathbb A}_{il} N_l^{(j)} + \varphi N_j^* \sum_{l \ne i} {\mathbb A}_{il} {\mathbb A}_{li} \, .
\end{equation}
This latter expression is the basis of the cavity treatment. I first
take an RMT average over the interaction matrix using Eq.~\eqref{randommatr}. The abundance of the removed $j^{\rm th}$ species is then
\begin{equation}\label{cav1}
N_j^*  = k_j - \mu \langle N_l^{(j)} \rangle + \gamma \sigma^2 \varphi N_j^* \, . 
\end{equation}
Differentiating with respect to $k_j$ and rearranging terms, this leads to a quadratic equation for the mean-field susceptibility,
\begin{equation}\label{cav3}
\varphi = 1 + \gamma \sigma^2 \varphi^2  \;\;\; \longrightarrow \;\;\;\; \varphi = \frac{1-\sqrt{1-4 \gamma \sigma^2}}{2 \gamma \sigma^2} \, , 
\end{equation}
where the chosen solution satisfies $\varphi \rightarrow 1$ when $\sigma \rightarrow 0$ as it should. It is remarkable that the generating function $C(\gamma \sigma^2)$ for Catalan numbers emerges here, 
as it did for the resummation of the perturbation 
expansion in Eq.~\eqref{expandN3}.
Finally taking average over abundances and intrinsic carrying capacities
in Eq.~\eqref{cav1} and assuming that abundance averages are the same for the full and the cavity system, $\overline N = \langle N_i^{(j)} \rangle$,
I finally obtain,
\begin{equation}\label{cav2}
\overline N = \frac{\kappa \varphi}{1 + \mu \varphi} = \kappa \frac{1-\sqrt{1-4 \gamma \sigma^2}}{2 \gamma \sigma^2 + \mu (1-\sqrt{1-4 \gamma \sigma^2})} \, .
\end{equation}
The cavity approach leads to a simple $\mu$-dependence in the form of an additional term $\mu \varphi$ in the denominator. Eq.~\eqref{cav2}  gives exactly the same result as the perturbative approach for $\mu=0$, however, for 
$\mu > 0$ the two approaches give functionally different results. I will discuss this discrepancy further in Section~\ref{sec:compare}.


\subsection{Variance of the Abundance Distribution}\label{sec:var_cav}


The above calculation of $\overline N$  first performed an average over the distribution of matrix elements ${\mathbb A}_{il}$ using 
Eq.~\eqref{randommatr2}. This procedure neglects fluctuations arising from the second term on the right-hand side of Eq.~\eqref{cav0} which 
need to be taken into account when calculating higher moments of the abundance distribution. Instead of Eq.~\eqref{cav1} I therefore
write
\begin{equation}\label{cav1_rep}
\frac{N_j^*}{\varphi}= k_j - \mu \langle N_l^{(j)} \rangle - \sum_{l \ne j} \delta {\mathbb A}_{jl} N_l^{(j)} \, ,
\end{equation}
where I additionally used Eq.~\eqref{cav3} to write $1-\gamma \sigma^2 \varphi=1-(\varphi-1)/\varphi=1/\varphi$.
With Eq.~\eqref{randommatr2},
the variance satisfies
\begin{equation}\label{cav1_rep}
\frac{\Sigma_N}{\varphi^2}= \chi^2 + \sigma^2 \langle (N_l^{(j)})^2 \rangle \, .
\end{equation}
With $\langle (N_l^{(j)})^2 \rangle = \Sigma_N + \overline{N}^2$ and Eq.~\eqref{cav3}, I finally obtain
\begin{eqnarray}\label{sigmu}
\Sigma_N&=&  \frac{(\chi^2 + \sigma^2 \overline N^2) \, (1-\sqrt{1-4 \gamma \sigma^2})^2}{4 \gamma^2 \sigma^4 - \sigma^2 (1-\sqrt{1-4 \gamma \sigma^2})^2} = \frac{(\chi^2+\sigma^2 \overline{N}^2)  \varphi^2}{1-\sigma^2 \varphi^2} \, ,
\end{eqnarray}
with the susceptibility $\varphi$ given in Eq.~\eqref{cav3}.
The $\mu$-dependence of $\Sigma_N$ is  contained only in $\overline N$, see Eq.~\eqref{cav2}.

\subsection{Single Parameter Scaling of Extinctions}\label{sec:scaling}

The abundance distribution is Gaussian. It is therefore uniquely determined by its mean and variance, and its truncation to $N_i^* \ge 0$ 
gives an excellent approximation to the true abundance distribution as long as the number of extinct species is small against the total number of species, $N_e/S \ll1$.\cite{Jac26} 
Ref.~\cite{Jac26} gave the
probability ${\cal Q}(N_e)$ to have $N_e$ extinct species as 
\begin{eqnarray}\label{prob1}
{\cal Q}(N_e) = 
\left(\begin{array}{c}
S \\ N_e
\end{array} \right) \,
F_-^{N_e} \, F_+^{S-N_e} \, , \:\;\;\;\;\;\;\;\;\;\;\;\;
F_{\pm}=\left(1 \pm {\rm erf}\left[\overline N/\sqrt{2 \Sigma_N}\right]\right)/2 \, ,
\end{eqnarray}
with the error function erf[x]. For fixed $S$, ${\cal Q}(N_e)$ depends on a single parameter $\eta :=\overline N/\sqrt{2 \Sigma_N}$. Following that observation, I conjectured in Ref.~\cite{Jac26} 
that the fraction $N_e/S$ of extinct species follows a single-parameter scaling. This conjecture was confirmed numerically, however, the single parameter could only be
assessed numerically, or analytically expressed in an approximate perturbative expression. In particular, the single parameter was observed
numerically to be $\mu-$independent for $\chi = 0$, which did not evidently emerge from the analytical expressions.

The cavity results of Eqs.~\eqref{cav2} and \eqref{sigmu}  instead deliver a remarkably concise analytical expression,
\begin{equation}
\eta^2 = \frac{\overline N^2}{\chi^2 + \sigma^2 \overline N^2} \frac{1-\sigma^2 \varphi^2}{2 \varphi^2} \, , \;\;\;\; {\rm with} \;\;\;\;\;  \overline N = \frac{\kappa \varphi}{1+\mu \varphi}  \;\;\; {\rm and} \;\;\; \varphi = \frac{1-\sqrt{1-4 \gamma \sigma^2}}{2 \gamma \sigma^2} \, .
\end{equation}
In particular when $\chi=0$, I obtain the simple expression,
\begin{equation}
\eta^2(\chi=0) = \frac{1-\sigma^2 \varphi^2}{2 \sigma^2 \varphi^2} \, ,
\end{equation}
which establishes the independence of $\eta$ on $\mu$ when $\chi=0$.

\subsection{Comparing Perturbative and Cavity Methods}\label{sec:compare}
For $\mu=0$, Eq.~\eqref{cav2} reproduces Eq.~\eqref{expandN3} -- the cavity and the resummed perturbative expansion give the exact same result. However, this 
ceases to be true when $\mu \ne 0$. Taylor expanding Eq.~\eqref{expandN3} in powers of $\gamma \sigma^2$ gives
\begin{equation}\label{taylor_expandN}
\overline N_{\rm pert} = \kappa \left[(1+\mu)^{-1} +(1+\mu)^{-2}\gamma \sigma^2+\left[(1+\mu)^{-2}+(1+\mu)^{-3}\right] \gamma^2 \sigma^4  + \frac{\mu^2+5 \mu+5}{(1+\mu)^4}\gamma^3 \sigma^6\right] + {\cal O}(\gamma^4 \sigma^8,S^{-1})\, ,
\end{equation}
while Taylor-expanding Eq.~\eqref{cav2} gives
\begin{equation}\label{taylor_cav}
\overline N_{\rm cav} = \kappa \left[(1+\mu)^{-1} +(1+\mu)^{-2}\gamma \sigma^2+\left[(1+\mu)^{-2}+(1+\mu)^{-3}\right] \gamma^2 \sigma^4  + \frac{2 \mu^2+6 \mu+5}{(1+\mu)^4}\gamma^3 \sigma^6\right] + {\cal O}(\gamma^4 \sigma^8,S^{-1})\, .
\end{equation}
Cavity and perturbative approaches start to differ at order $\gamma^3 \sigma^6$ with a difference $\langle N_i^* \rangle_{\rm pert} - \langle N_i^* \rangle_{\rm cav}=-\kappa \, \mu \, \gamma^3 \sigma^6/(1+\mu)^3 + {\cal O}(\gamma^4 \sigma^8,S^{-1})$, whose value is maximal for $\mu=0.5$. I obtain relative
differences $[\overline N_{\rm pert} - \overline N_{\rm cav}]/\overline N_{\rm cav} \le 0.025$ 
in the regime of validity of the perturbative approach, $\sigma (1 +|\gamma|) < 1$.

In Section~\ref{sec:nums} I investigate numerically how and when the two methods differ for both $\overline N$ and $\Sigma_N$.

\section{Sparse Systems}\label{sec:sparse}

Ecological networks significantly differ from the all-to-all interaction networks considered so far.
They are instead characterized by structured, sparse interaction networks that do not follow the random matrix assumption of  Eq.~\eqref{randommatr}.~\cite{Pim91,Wil00,Kra03,The10,Del19} Generically, most interspecies interaction links  vanish, $\delta {\mathbb A}_{ij}=0$, as 
the presence of species $i$ does not affect most of the other $S-1$ species $j \ne i$. The average number $L$ of nonzero interaction links in $\delta {\mathbb A}$
defines the sparsity $\zeta = L/S^2$ of the interaction network. In ecological networks, one typically has $\zeta \in [0.05,0.4]$,~\cite{Wil00,Del19} and
the underlying interaction network is very different from the full RMT model. 
A related question is whether the same sparsity affects the average, non-fluctuating part ${\mathbb M}$ of the interaction matrix. One may think, for instance that 
the presence of a species influences other species, simply because "it is there" and contributes an interspecies logistic term to Eq.~\eqref{lv}. I therefore introduce two sparsities, $\zeta_\mu$ and $\zeta_\sigma$, and modify Eq.~\eqref{lv} as
\begin{subequations}
\label{randommatr_sp}
\begin{eqnarray} 
\langle {\mathbb A}_{ij}\rangle & = & \zeta_\mu \mu/S \;, \\
\label{xcorr}
\langle {\mathbb A}_{ij} {\mathbb A}_{kl} \rangle & = & \zeta_\sigma \sigma^2 ( \delta_{ik} \delta_{jl}+\gamma \delta_{il} \delta_{jk})/S + (\zeta_\mu \, \mu/S)^2 \, .
\end{eqnarray}
\end{subequations}
The resulting model has the advantage that both the cavity and perturbative approaches followed above can still be applied. One straightforwardly 
obtains the same results as in Eq.~\eqref{expandN5}, \eqref{sigma2}, \eqref{cav2} and \eqref{sigmu}, however
with the substitutions 
\begin{equation}\label{sparse}
\sigma \rightarrow \zeta_\sigma^{1/2} \sigma \;\;\; {\rm and} \;\;\; \mu \rightarrow \zeta_\mu  \mu \; .
\end{equation}
Note the scaling with $\zeta_\sigma^{-1/2}$ which comes about from the assumption that if $\delta {\mathbb A}_{ij} \ne 0$, then $\delta {\mathbb A}_{ji} \ne 0$ -- if species $i$ influences one way or another
species $j$, then the reverse is true. 

\section{Numerical Simulations}\label{sec:nums}

I illustrate the validity of the above analytical results through numerical computations of $\overline N$ and $\Sigma_N$ for an ensemble of RMT realizations of Eq.~\eqref{equi2} with $S=607$.
Two different analytical approaches gave two different results for the lowest moments $\overline N$ and $\Sigma_N$ of the abundance distribution when $\mu > 0$. The perturbative method is based on 
a controlled expansion valid for $\sigma < 1/(1+|\gamma|)$. The cavity approach has a larger regime of validity $ \sigma < 1/2 \sqrt{| \gamma|}$ because $|\gamma| \le 1$. It is
furthermore based on a mean-field approximation which usually is valid in sufficiently connected, high-dimensional systems.~\cite{chandler}
In our case, this translates into $S \gg 1$ and $\zeta$ not too small. Additionally, Eqs.~\eqref{cav2} and \eqref{sigmu} require $\sigma < 1/2 \sqrt{|\gamma|}$. 
A priori, the two methods do not have the same regime of validity, and provided $S \gg 1$ and the interaction matrix is not too sparse (to justify a mean-field approximation)
one expects the cavity method to have a larger regime of validity. 

\begin{figure}[h]
\center
\includegraphics[width=0.4 \columnwidth]{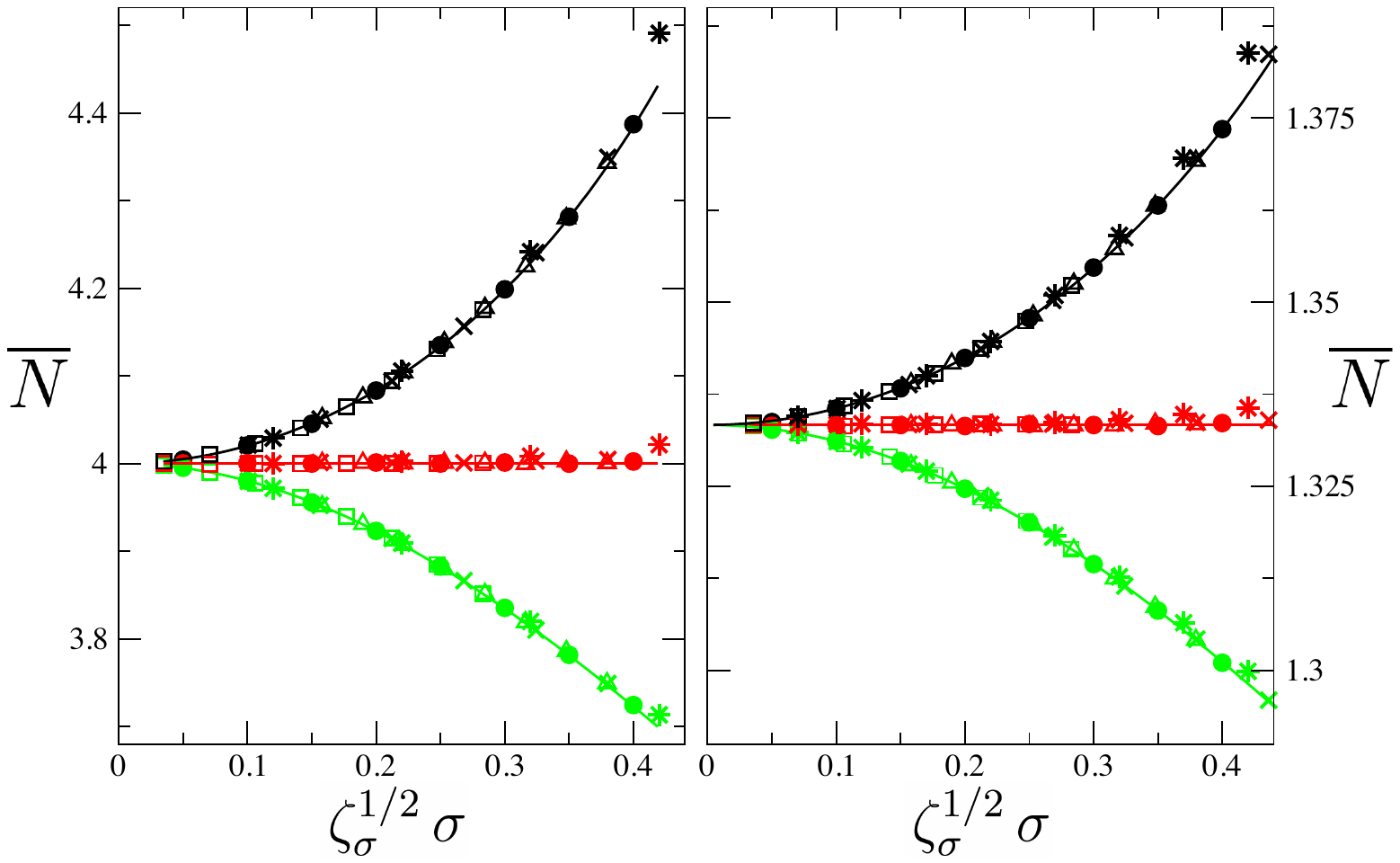}
\includegraphics[width=0.4 \columnwidth]{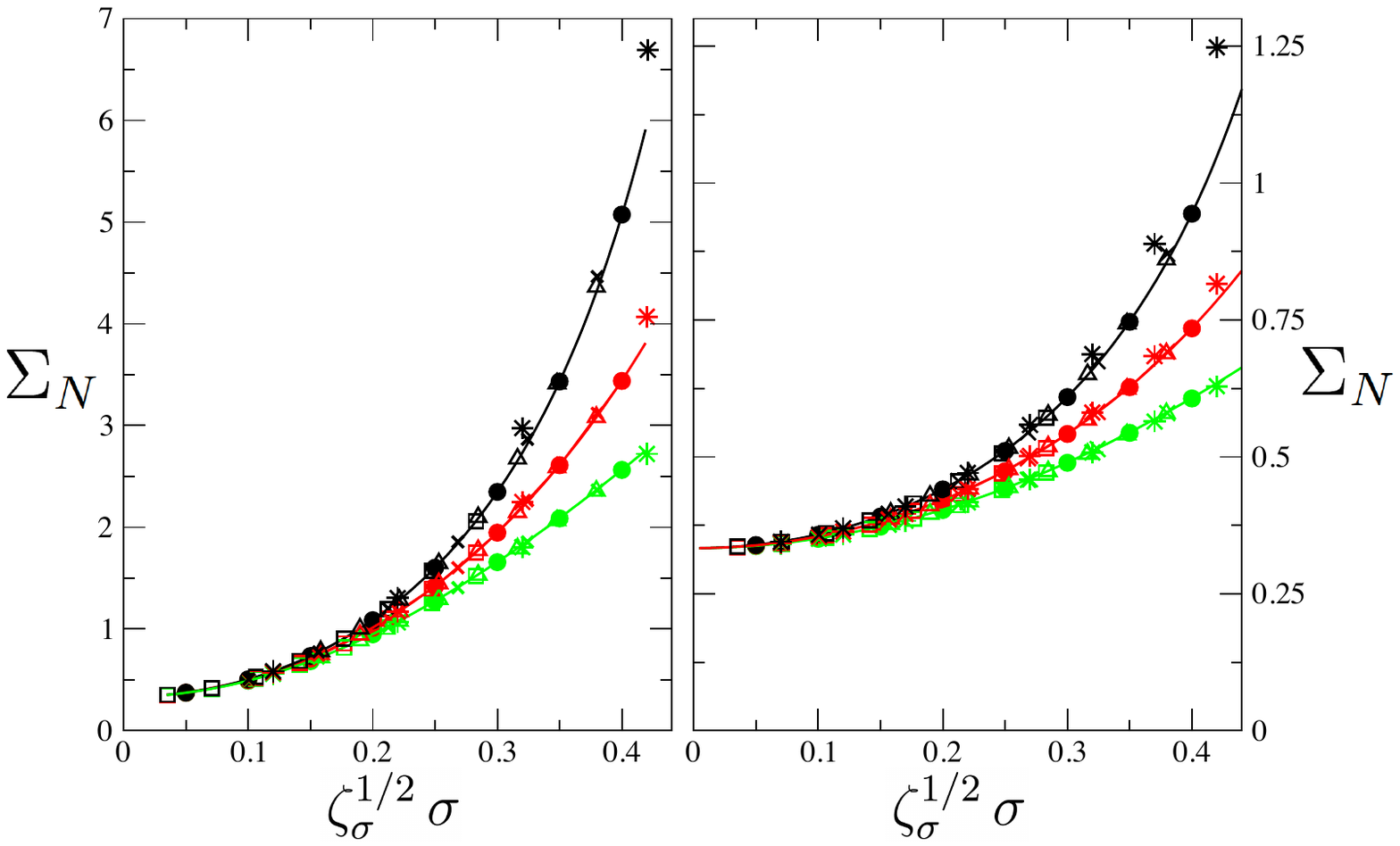}
\vspace{-3mm}
\caption{Left panel: Average of the abundance distribution at stable equilibria for Eq.~\eqref{lv}, 
$\mu=0$ (left) and $\mu=2$ (right), and a random, homogeneous distribution $k_i \in [3,5]$, i.e. with $\kappa=4$ and $\chi=1/\sqrt{3}$,
and for $\gamma=-0.5$ (green), 0 (red) and 0.5 (black). 
Right panel: Variance of the abundance distribution for the same cases. 
Different sparsity of $\zeta_\sigma =1$ (full circle) $\zeta_\sigma=0.5$ (squares), $\zeta_\sigma=0.1$ (triangles), $\zeta_\sigma=0.05$ ($\times$) 
and $\zeta_\sigma=0.01$ ($*$) 
are displayed. In all cases, the average, logistic interaction is not sparse, $\zeta_\mu=1$.
Solid lines give the RMT prediction, Eq.~\eqref{expandN5}, and  the  cavity prediction, Eq.~\eqref{cav2}. They are indistinguishable in this regime.
Data correspond to averages over all species for 500 realizations of the random interaction matrix $\mathbb A$ with $S=607$.
\label{fig6}  }
\end{figure}

There is a catch, however, that we are not interested in the properties of the 
fixed-point of Eq.~\eqref{equi2}, instead, we want to understand the statistics of population abundances for asymptotic stationary solutions
to Eq.~\eqref{lv}, i.e. for feasible fixed-points of Eq.~\eqref{equi2}, with $N_i^* \ge 0$, $\forall i$. Ref.~\cite{Jac26} showed that the abundance of 
feasible fixed points differs from those of solutions to Eq.~\eqref{equi2} only by shifts $\delta N_i^* = {\cal O}(N_e/S)$ that remain small as long as the fraction of extinct species is small. 
Therefore, I first present here numerical investigations restricted to parameter values with less than 5\% of extinct species on average.

The left panel of Fig.~\ref{fig6} shows $\overline N$ vs. $\sigma$ for different values of $\gamma$, $\mu$ and sparsity $\zeta_\sigma$. 
In all cases, the analytical predictions almost exactly match the numerical data. Furthermore, in
that parameter range, there is no noticeable difference between the perturbative and the cavity results -- the two analytical curves lie on top of one another and are undistinguishable. 
The right panel of Fig.~\ref{fig6} shows that the same conclusion applies to $\Sigma_N$.
Both panels show data with sparsity $\zeta_\sigma=1$, 0.5, 0.1, 0.05 and 0.01 and $\zeta_\mu=1$. The scaling with $\zeta_\sigma^{1/2} \sigma$ predicted by Eq.~\eqref{sparse}  is
confirmed, at least down to $\zeta_\sigma=0.05$, with small, but nonegligible deviations occurring for $\gamma=0.5$, $\zeta_\sigma=0.01$ and the largest values of $\sigma$.
This shows that our approach is valid in the reported regime of sparsity of real ecosystems, $\zeta \in [0.05,0.4]$,~\cite{Wil00,Del19} but may fail for very weakly connected 
interaction networks.
Note that the choice $\zeta_\mu=1$ has been made to better illustrate the scaling with $\zeta_\sigma$ predicted by Eq.~\eqref{sparse} without resorting to too many figures. 
In all the cases shown, the ratio of extinctions is always below 5 \% for all but the rightmost datapoint for each 
sparsity value, for which some RMT realisations have between 5 and 10 \% of extinct species.

\begin{figure}[h]
\center
\includegraphics[width=0.45 \columnwidth]{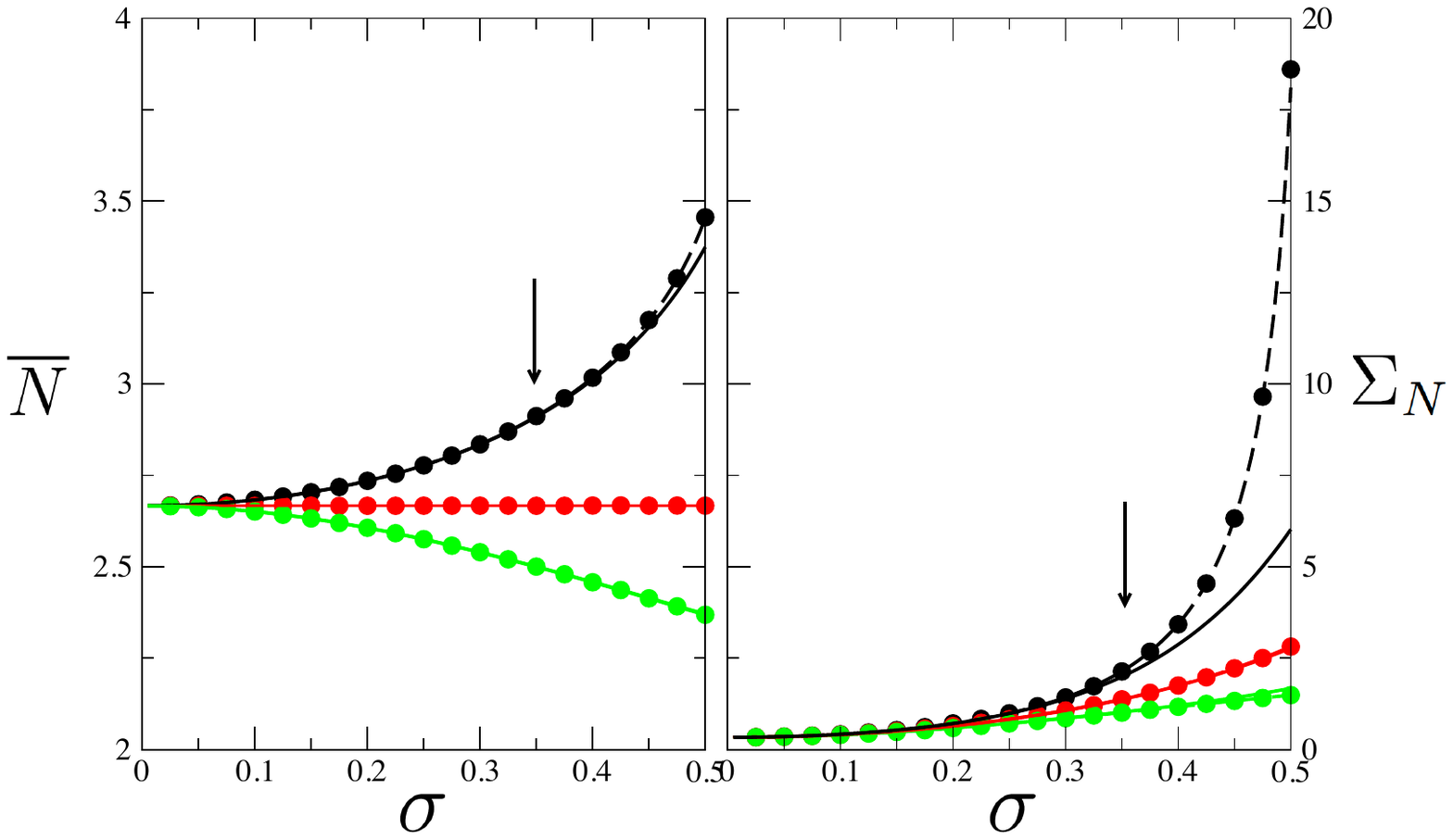}
\vspace{-3mm}
\caption{Average (left) and variance (right) of the abundance distribution at stable equilibria for Eq.~\eqref{lv}, 
$\mu=0.5$, a distribution of $k_i $ with $\kappa=4$ and $\chi=1/\sqrt{3}$,
as a function of $\sigma$ and for $\gamma=-0.9$ (green), 0 (red) and 0.9 (black). 
Dashed lines give the cavity prediction and solid lines the RMT expansion prediction.
The arrows show $\sigma=0.35$, where each RMT realization gives at most 5 \% of extinctions; higher values of $\sigma$ give higher ratio of extinctions, up to 35 \% at $\sigma=0.5$.
Data correspond to averages over all species for 500 realizations of the random interaction matrix $\mathbb A$ with $S=607$.\\[-6mm]
\label{fig7}  }
\end{figure}

In the regime $N_e/S \ll 1$ of interest for population dynamics, both theories work very well and give very accurate values for $\overline N$ and $\Sigma_N$. 
Despite them giving functionally different results -- see Eq.~\eqref{expandN5} and \eqref{cav2} for $\overline N$, and Eq.~\eqref{sigma2} and \eqref{sigmu} for $\Sigma_N$ -- 
they are indistinguishable as long as $N_e/S \ll 1$. One may still wonder which of the two methods gives the right answer in their regime of validity, even when there is
a significant fraction of extinct species. After all, finding the distribution of components of $\vec{N^*}$ in Eq.~\eqref{equi2} is a mathematical problem on its own. 
To try and answer this question I set $\mu=0.5$ -- this  maximizes the difference between $\overline N_{\rm pert}$ and $\overline N_{\rm cav}$ in Eqs.~(\ref{taylor_expandN},\ref{taylor_cav}) -- $\gamma=0.9$ and consider interaction strengths  up to $\sigma=0.5$, which pushes the system closer to the edge of the regime of validity for both approaches.  
The results are shown in Fig.~\ref{fig7}.
Data for $\overline N$ in the left panel are still barely distinguishable, until $\sigma \approx 0.4$, above which, the cavity result (dashed line) still follows the numerical data, while
the resummed perturbative result starts to deviate  from it slightly. The deviation is not large, yet significant, which indicates that the cavity approach gives the right answer. This conclusion is further
corroborated by results for $\Sigma_N$ in the right panel of Fig.~\ref{fig7}. Keep  in mind however that the perturbative approach is truncated at order $\sigma^6$ 
for $\Sigma_N$, and that at least  part of the larger deviation of the perturbative result from the numerical data comes from neglected higher orders. 

To conclude this numerical section I note that, regardless of the number of negative abundance species, all fixed points have been found to be stable. Feasibility is generically lost before stability.

\section{Discussion}\label{conclusion}

Both methods developed above give a remarkable result for $\overline N$ as a closed-form formula. While the cavity approach also delivers a closed-form formula for $\Sigma_N$, 
I have been unable to resum the corresponding infinite series for the perturbative approach. 
Nevertheless, two major steps forward compared to Ref.~\cite{Jac26} are that (i) for each order, the exact $\mu$-dependence can be calculated, and (ii) the expansion can be 
extended to any order in $\sigma$ thanks to clearly expressed diagrammatic Feynman rules.

Rather interestingly, the two approaches give functionally different closed-form expressions for $\overline N$, except when $\mu=0$. 
The two expressions match numerical data and are furthermore undistinguishable from one another most of the time. However, pushing parameters to the boundary of their
regime of validity, the perturbative approach starts deviating from numerical data, while the cavity approach keeps reproducing numerical data almost exactly.  
This is rather surprising: the cavity approach is based on a mean-field approximation which, while making sense, is not justified {\it per se}. On the other hand the approximations
on which the perturbative approach is based seem to be well-controlled. Why the cavity approach performs better than the perturbative approach is an open question. Perhaps
 subdominant diagrams neglected in the RMT perturbative approach become larger and larger as one reaches the boundary of validity of the theory. I have not been able to 
 figure that out conclusively. 

A conclusion that was already drawn in Ref.~\cite{Jac26} is that, with the scaling of the interaction strength with $S$ as in Eq.~\eqref{lv}, feasibility is lost before stability: even when $\overline N \gg \sigma$,
there will be a critical number species $S_c$ above which more and more species populate one negative hyperoctant of the abundance hyperspace (with at least one negative abundance). Using standard approximation for error functions
I get a parametric relation $S \simeq \exp[-4 \sigma^2]$, which either gives a critical number of species 
at fixed $\sigma$, $S_c(\sigma)\simeq \exp[-4 \sigma^2]$, or 
a critical interaction strength at fixed $S$, $\sigma_c(S) \simeq 1/\sqrt{2 \ln S}$, above which some  species become extinct. This latter estimate agrees with Refs.~\cite{Dou18,Biz21}

\section{Use of AI}\label{sec:ai}

This manuscript has been entirely written by my human, natural intelligence, which furthermore drew all figures, wrote the numerical codes 
and analyzed the numerical data. There was no editorial help from AI at any stage. 

In my earlier work, Ref.~\cite{Jac26} I constructed a new analytical method to the problem of feasibility and species extinctions in generalized Lotka-Volterra models. Thanks to that method
I obtained a perturbative series for the moments of the distribution of abundances that was fully consistent up to and including the order $\sigma^4$. However, I could  calculate 
the $\sigma^6$ correction only for $\mu=0$. This was not entirely satisfying and I asked several AI chatbot to read Ref.~\cite{Jac26} and help me obtain the $\sigma^6$ term for $\mu \ne 0$.
I used only free versions and
three of them could "understand" the mathematical framework -- GPT-5.6 Luna, Claude Sonnet 5.5 and Gemini 3.6 Flash. The corresponding chats are publicly available.~\cite{claude,gemini,chatgpt}

Gemini "understood" that the perturbative expansion of Ref.~\cite{Jac26} gives ${\overline N}$
as a series of non-crossing planar graphs. I knew this at the time, however this was not in Ref.~\cite{Jac26} nor did I fed Gemini that information in our chat. Gemini 
further gave the combinatorial connection to Catalan and Narayana numbers and got to the 
closed-form expressions of Eq.~\eqref{expandN3} and \eqref{expandN5} via their generating functions -- though I had to correct it several times.
At the time, I had only realized the connection to non-crossing planar graphs,
not  the one to Catalan and Narayana numbers, which led to the exact resummation of the infinite expansion for $\overline N$, Eq.~\eqref{expandN5}.

Claude spontaneously suggested to instead use the cavity method to calculate $\overline N$ and $\Sigma_N$.~\cite{caveat2} Both Gemini and Claude made very interesting 
connections to domains of research I was unfamiliar with -- Gemini to the combinatorics of planar graphs, with Catalan and Narayana numbers, as well as Dyck paths; Claude to the cavity approach to complex systems. 

From a calculational point of view, however, both Gemini and Claude had to be closely supervised at all times, as they seem to be unable to figure out on their own whether their result is right or wrong -- I had to correct by hand
(with paper and pen) prefactors, minus signs and other computational mistakes, many of which were  easily spotted by my human intelligence. Finally, it was interesting to see that Gemini, when confronted 
with different expressions for the perturbative expansion of $\Sigma_N$ was not able to figure out which one was correct.~\cite{gemini} In the end, I got very little help, if any, from AI chatbots 
for the perturbative approach to $\Sigma_N$. I did almost all 
the perturbative calculation for $\Sigma_N$ on my own, as that problem seemed to exceed what GPT-5.6 Luna, Claude Sonnet 5 and Gemini 3.6 Flash could do. 

Finally, GPT-5.6 Luna seemed to be converging toward similar conclusions as Gemini and Claude, though more slowly. It however circumvoluted, perhaps 
attempting to motivate me to upgrade to a paying version. Not willing to do that yet, I did not get anything really useful out of GPT-5.6 Luna.~\cite{chatgpt}

Finally, I tried to extract information on the human-generated knowledge behind each chatbot's answer, because scientific credit must go there and nowhere else. 
While I checked all obtained references for potential AI hallucination, I bear full responsibility for any error in citation or missing reference. \\

\end{document}